\documentclass[11pt]{article}
\usepackage{amssymb}
\usepackage{amsfonts}
\usepackage{amsthm}
\usepackage{latexsym}
\usepackage{gensymb}
\usepackage[cmex10]{amsmath}
\usepackage[mathscr]{eucal}
\usepackage{algpseudocode}
\usepackage{algorithm}
\usepackage{subfigure}
\usepackage{subfigmat}
\usepackage{array}
\usepackage{psfrag}
\usepackage{graphicx}
\usepackage{color}
\usepackage{xcolor}
\usepackage{cite}
\usepackage[bottom]{footmisc}
\usepackage{epstopdf}
\usepackage{authblk}
\title{Advanced dispersion engineering of a III-Nitride micro-resonator for a blue/UV frequency comb}
\author[1,2]{Ali Eshaghian Dorche\footnote{This work was performed while A. E. D. was a 2019 summer intern at PARC.}}
\author[2]{Do{\u g}an Timu{\c c}in\footnote{Corresponding author contact:  \texttt{dtimucin@parc.com}}}
\author[2]{Krishnan Thyagarajan}
\author[2]{Thomas Wunderer}
\author[2]{Noble Johnson}
\author[2]{David Schwartz}

\affil[1]{School of Electrical and Computer Engineering, Georgia Institute of Technology, \break 777 Atlantic Drive NW, Atlanta, GA 30332, USA}
\affil[2]{Palo Alto Research Center (PARC), A Xerox Company, \break 3333 Coyote Hill Road, Palo Alto, CA 94304, USA}
\date{\today}

\begin{document}

\maketitle

\begin{abstract}
A systematic dispersion engineering approach is presented toward designing a III-Nitride micro-resonator for a blue/UV frequency
comb.  The motivation for this endeavor is to fill the need for compact, coherent, multi-wavelength photon sources that can be paired with, \emph{e.g.}, the $^{171}\textrm{\textrm{Yb}}^{+}$ ion in a photonic integrated chip for optical sensing, time-keeping, and quantum computing applications.  The challenge is to overcome the normal material dispersion exhibited by the otherwise ideal (\emph{i.e.}, low-loss and large-Kerr-coefficient) AlGaN family of materials, as this is a prerequisite for bright-soliton Kerr comb generation.  The proposed approach exploits the avoided-crossing phenomenon in coupled waveguides to achieve strong anomalous dispersion in a desired wavelength range.  The resulting designs reveal a wide range of dispersion response tunability, and are realizable with the current state-of-the-art growth and fabrication methods for AlGaN semiconductors.  Numerical simulations of the spatio-temporal evolution of the intra-cavity field under continuous-wave laser pumping indicate that such a structure is capable of generating a broadband blue/UV bright-soliton Kerr frequency comb.
\end{abstract}

\section{Introduction}
Frequency standards play an important role both in fundamental science and in a wide range of applications such as sensing, metrology, and position, navigation \& timing, to name just a few.  The field was revolutionized by the advent of the optical frequency comb  \cite{Hans2006}.  Over the past decade, integrated photonic approaches to optical frequency comb generation have received much attention as a potential technology for enabling ultra-precise measurements in a miniaturized platform \cite{DSA$^{+}$2007, KHD2011}, wherein the generation of stable, coherent frequency components (also called comb lines) is achieved through the injection of a continuous-wave (CW) laser into a micro-resonator exhibiting third-order (\emph{i.e.}, Kerr) nonlinearity \cite{optical comb, Chem2016}.  A dissipative Kerr soliton provides a tool for generating coherent and low-noise comb lines over a wide spectral range, circumventing the challenges associated with sub-comb processes in non-solitonic Kerr combs, which suffer from much higher noise.

There has been significant progress on both theoretical and experimental fronts, to the point where we now have fully developed spectro- and spatio-temporal models \cite{chembo-cmt, chembo-lle} of the nonlinear dynamical Kerr-comb generation process inside a micro-resonator, advanced dispersion engineering approaches \cite{ali1, ali2, ali3, A.Weiner1} for micro-resonator design optimization, as well as fabrication processes for Si$_{3}$N$_{4}$-based structures \cite{lipsonHQ, A.WeinerHQ, TJKippenberg_Damascene}, allowing the manufacture of near-IR micro-resonators with very high quality factors ($Q$) in a variety of material platforms (Si, SiO$_{2}$, Si$_{3}$N$_{4}$, SiC, diamond, AlN, AlGaAs, GaP, MgF$_{2}$).  These advances have enabled the laboratory demonstrations of frequency-bin entangled photon sources and Doppler-cooling of atoms/ions \cite{combcooling, frequencybin1}, among others.

While there has been extensive reporting on efficient bright-soliton Kerr-comb generation at mid-infrared, and more recently at near-visible, wavelengths as highlighted above, there has been no report of direct Kerr-comb generation in the blue and UV end of the spectrum.  This is largely due to the significant normal dispersion exhibited by candidate materials that otherwise have desirable features such as large bandgaps and reasonably strong third-order nonlinearity at these wavelengths.  Indeed, ternary and quaternary III-V semiconductors, and specifically the group III-N materials, are very promising for ultrashort pulse generation at short wavelengths, but are unfortunately normally dispersive.  

Anomalous dispersion is a favorable, if not essential, ingredient for generating broadband bright-soliton Kerr combs.  This is because soliton formation is greatly aided by the anomalous dispersion of the cold (unloaded) cavity, which can compensate for the (normal) Kerr-nonlinearity dispersion of the loaded resonator.  Purely geometric approaches to achieving anomalous dispersion typically lead to feature sizes that are impractical to fabricate in the blue/UV wavelength regime, therefore necessitating a more advanced dispersion engineering approach.

In this paper, we describe a systematic dispersion engineering approach for achieving anomalous dispersion in an AlGaN micro-resonator at blue/UV wavelengths.  At the core of the approach is the avoided-crossing behavior exhibited by coupled weaveguides, which we use to our advantage in designing a hybrid structure that is at once reasonably easy to fabricate and highly tailorable in its dispersion response.  Similar ideas have been used before at visible and near-IR wavelengths, working with Si or Si$_{3}$N$_{4}$ where $Q$ can be very high \cite{AlanW1, alanW2, ZJ1}; the added challenge here is to achieve anomalous dispersion in a micro-resonator with a moderate $Q$.  We then describe an AlGaN heterostructure that embodies these dispersion engineering principles toward
the demonstration of a Kerr comb with our micro-resonator design.

\begin{figure}[th!]
\centering
\includegraphics[width = 0.75\linewidth]{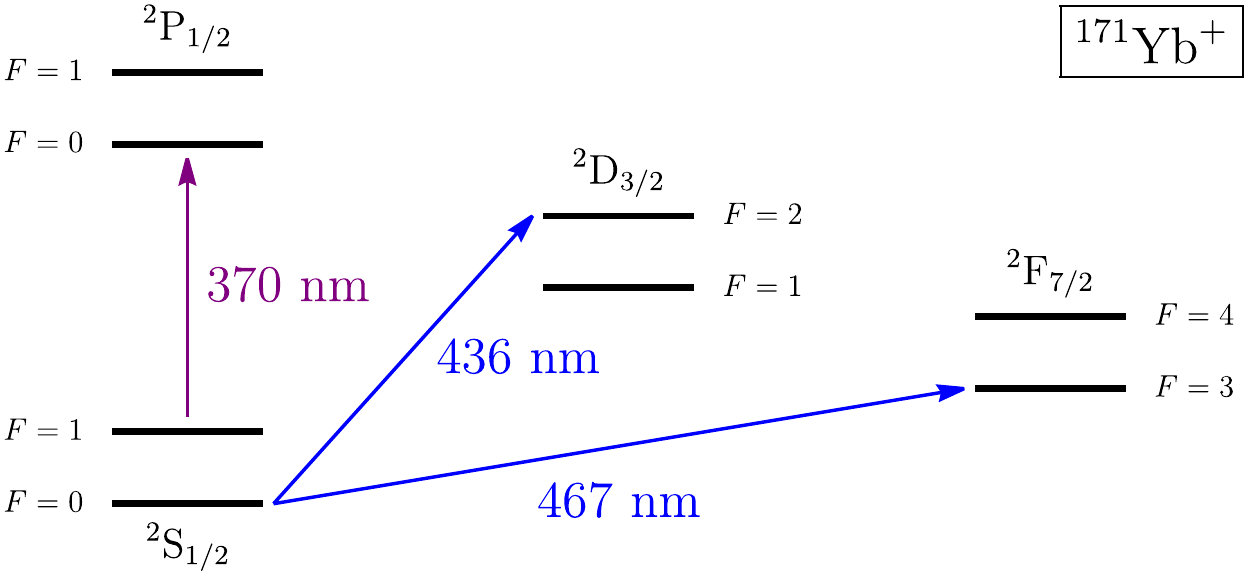}
\caption{Energy diagram for the $^{171}{\textrm{Yb}}^{+}$ ion (adapted from \cite{CLK$^{+}$2017}), showing the blue and UV transitions germane to our discussion:  $436$~nm and $467$~nm optical clock transitions (blue arrows) and $370$~nm Doppler-cooling transition (purple arrow).}
\label{fig:Yb}
\end{figure}

The utility of an AlGaN-based blue/UV frequency comb design is evidenced by the recent interest in optical timing, trapping, and quantum computing applications employing Ytterbium, which has a number of suitable optical transitions; see Figure~\ref{fig:Yb}.  The application of AlN for optical waveguides and micro-resonators has been explored recently \cite{LFT$^{+}$2018, LBG$^{+}$2018}.  Pumped by a quantum-well UV/visible laser grown and fabricated on a common nitride epitaxial wafer \cite{WCN$^{+}$2012}, a micro-resonator generating a broadband Kerr comb would be an ideal integrated photonic source for such applications.

\section{Proposed method}
An individual waveguide mode with a propagation constant $\beta(\omega)$ has a group velocity $1/v_{\textrm{g}} = \mathrm{d}\beta/\mathrm{d}\omega \equiv \beta'(\omega)$, whose (frequency) dispersion is, in turn, given by $\mathrm{d}^{2}\beta/\mathrm{d}\omega^{2} \equiv \beta''(\omega)$.  The two group velocity dispersion (GVD) regimes are delineated by the sign of this latter quantity; \emph{i.e.}, $\beta'' > 0$ for normal, and $\beta'' < 0$ for anomalous.  The III-Nitride materials exhibit strong normal dispersion in the blue/UV part of the spectrum, and therefore are not inherently suitable for bright-soliton generation.

Consider instead \emph{two} waveguides with intrinsic propagation constants $\beta_{1}$ and $\beta_{2}$ that exhibit coherent,  evanescent, co-directional coupling.  The mutual coupling strengths $\kappa_{12}$ and $\kappa_{21} = \kappa_{12}^{*}$ can be tailored by adjusting the thickness of the gap between the guides.  It is well known (see, \emph{e.g.}, \cite{Chua2009}, \S 8.2) that such a hybrid structure possesses a pair of super-modes with propagation constants
\begin{equation}
\beta_{\pm} = \bar{\beta} \pm \sqrt{\Delta\beta^{2} + \kappa^{2}},
\label{eq:betapm}
\end{equation}
where $\kappa = \sqrt{\kappa_{12} \kappa_{21}} = |\kappa_{12}|$, and we defined
\begin{eqnarray}
\bar{\beta} & = & \tfrac{1}{2} \, (\beta_{1} + \beta_{2}),
\label{eq:betabar} \\
\Delta\beta & = & \tfrac{1}{2} \, (\beta_{1} - \beta_{2}).
\label{eq:betadel}
\end{eqnarray}
These super-modes are further distinguished by their symmetric ($+$) and anti-symmetric ($-$) spatial field profiles over the transverse cross-section of the hybrid structure.

Now, let $\omega_{\textrm{c}}$ denote the central frequency around which we wish to engineer anomalous dispersion in the hybrid structure.  The geometric parameters of the high-index layers are then designed to achieve phase matching between the two guides at this frequency; \emph{i.e.},
\begin{equation}
\beta_{1}(\omega_{\textrm{c}}) = \beta_{2}(\omega_{\textrm{c}}).
\label{eq:betawc}
\end{equation}
The propagation constants $\beta_{\pm}(\omega)$ of the \emph{hybrid guide} display the so-called ``avoided crossing'' (also known as ``level repulsion'') behavior around $\omega_{\textrm{c}}$.  This effect, controlled through the coupling parameter $\kappa$, makes it possible to push $\beta_{-}''$ toward negative values, thus suppressing the normal material dispersion and achieving a net anomalous dispersion in the odd super-mode of the hetero-structure.

In order to explore this possibility in detail, we differentiate \eqref{eq:betapm} with respect to $\omega$ to obtain
\begin{eqnarray}
\beta_{\pm}' & = & \bar{\beta}' \pm \frac{\Delta\beta \, \Delta\beta' + \kappa \, \kappa'}{\sqrt{\Delta\beta^{2} + \kappa^{2}}},
\nonumber \\
\beta_{\pm}'' & = & \bar{\beta}'' \mp \frac{(\Delta\beta \, \Delta\beta' + \kappa \, \kappa')^{2}}{(\Delta\beta^{2} + \kappa^{2})^{3/2}} \pm \frac{(\Delta\beta')^{2} + \Delta\beta \, \Delta\beta'' + (\kappa')^{2} + \kappa \, \kappa''}{\sqrt{\Delta\beta^{2} + \kappa^{2}}}.
\label{eq:betapp}
\end{eqnarray}
Evaluating these expressions at $\omega = \omega_{\textrm{c}}$ by making use of \eqref{eq:betadel} and \eqref{eq:betawc}, we find
\begin{eqnarray}
\beta_{\pm}'(\omega_{\textrm{c}}) & = & \bar{\beta}'(\omega_{\textrm{c}}) \pm \kappa'(\omega_{\textrm{c}}),
\nonumber \\
\beta_{\pm}''(\omega_{\textrm{c}}) & = & \bar{\beta}''(\omega_{\textrm{c}}) \pm \left\{\frac{\left[\Delta\beta'(\omega_{\textrm{c}})\right]^{2}}{\kappa(\omega_{\textrm{c}})} + \kappa''(\omega_{\textrm{c}})\right\}.
\label{eq:betappwc}
\end{eqnarray}
The first term on the right-hand side of \eqref{eq:betappwc} is positive, since the individual guides are assumed at the outset to have normal dispersion.  Fortunately, however, the first term in the curly braces is also strictly positive, and if it can be made sufficiently large by a judicious choice of material and geometric parameters, then the whole right-hand side of \eqref{eq:betappwc} can be made to go negative for (and only for) the odd super-mode, thus rendering it anomalously dispersive, as desired.  This effect persists over a band of frequencies around $\omega_{\textrm{c}}$, beyond which the guides become phase-mismatched and the dispersion behavior reverts to the normal regime.

\section{Dispersion engineering}
We now choose a basic hybrid waveguide structure to demonstrate the approach proposed above.  We exclusively focus on the odd super-mode, and drop the subscript on $\beta_{-}$.  Defining an effective refractive index through $\beta = (\omega/c) \, n_{\textrm{eff}}$, the GVD parameter of this super-mode will be computed via
\begin{equation*}
D_{\lambda} = -\frac{\lambda}{c} \, \frac{\mathrm{d}^{2} n_{\textrm{eff}}}{\mathrm{d}\lambda^{2}},
\end{equation*}
where $c$ is the vacuum speed of light, and we switched to the customary wavelength variable $\lambda = 2 \pi c/\omega$, which makes it more convenient to work with  refractive-index data for optical materials.  The anomalous dispersion behavior will now be indicated by $D_{\lambda} > 0$.

\begin{figure}[t!]
\centering
\includegraphics[trim = 10 190 190 10, width = 1\linewidth]{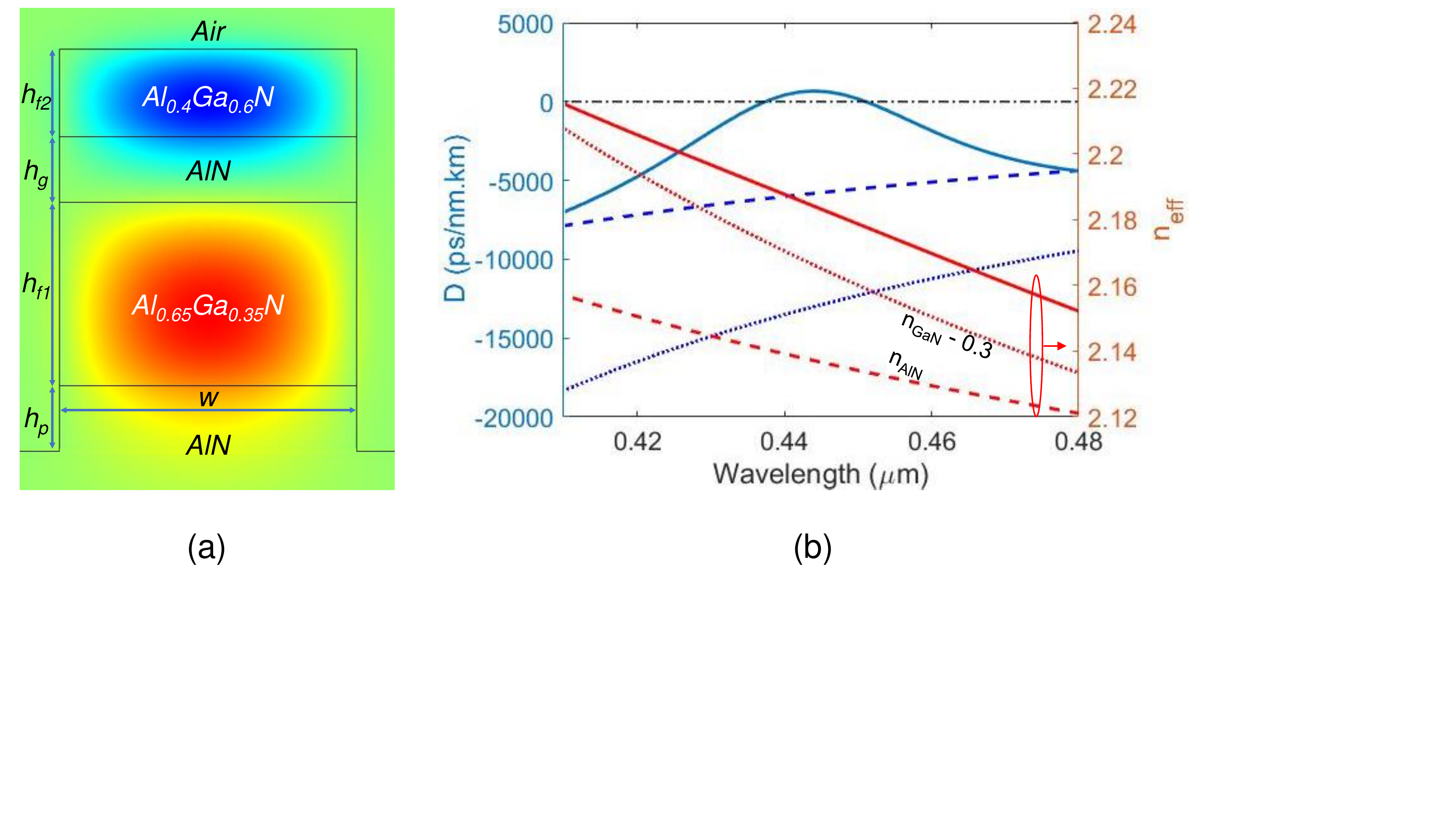}
\caption{(a) A representative AlGaN hybrid waveguide structure and the intensity profile of its fundamental odd super-mode at $\lambda = 442$~nm.  The dimensions are $h_{\textrm{p}} = 150$~nm, $h_{\textrm{f}1} = 440$~nm, $h_{\textrm{g}} = 250$~nm, $h_{\textrm{f}2} = 200$~nm, and $w = 700$~nm.  The Al mole fractions are $x_{1} = 0.65$ and $x_{2} = 0.4$.  (b) The GVD parameter $D_{\lambda}$ (solid blue) and the effective refractive index $n_{\textrm{eff}}$ (solid red) of the odd super-mode as functions of wavelength near the avoided-crossing point.  The dashed and dotted curves show the GVD parameters (blue) and the (ordinary) refractive indices (red) for \emph{bulk} AlN and GaN, respectively (data from \cite{SSPE2016}; the index of bulk GaN is shifted down by $0.3$ for plotting purposes).}
\label{fig:guide}
\end{figure}

Figure~\ref{fig:guide}(a) shows a vertical stack of two rectangular waveguides with heights $h_{\textrm{f}1}$ and $h_{\textrm{f}2}$ and a common width $w$, separated by a gap of height $h_{\textrm{g}}$ and sitting on a pedestal of height $h_{\textrm{p}}$.  The two guides are made of $\textrm{Al}_{x} \textrm{Ga}_{1 - x} \textrm{N}$ (with $x_{1} \neq x_{2}$, in general), whereas the gap and the pedestal are made of AlN.  The waveguide heterostructure is shown on an AlN substrate with no (dielectric) encapsulation.

We studied the eigenmodes of this structure numerically using the Wave Optics Module of the finite-element modeling software COMSOL Multiphysics.  With consideration of the blue/UV transitions of the $^{171}{\textrm{\textrm{Yb}}}^{+}$ ion in mind and guided by the analytical results of the previous section, we selected the geometric parameters and the material compositions of the structure such that the avoided-crossing wavelength falls around $442$~nm.  As shown in Fig.~2, the fundamental odd super-mode at this wavelength has an anti-symmetric transverse field profile, as expected, which is a perturbed superposition (with a phase shift of $\pi$~rad) of the two quasi-TE (HE$_{00}$) modes of the individual rectangular guides.  The figure caption lists the parameters of this particular structure, which are readily achievable via standard epitaxial growth techniques.

Figure~\ref{fig:guide}(b) shows the GVD parameter as well as the effective refractive index of the odd super-mode.  As the figure reveals, both AlN and GaN are normally dispersive at these wavelengths, and yet, this particular design of the hybrid structure is able to achieve anomalous dispersion near the target wavelength interval, as indicated by the portion of the blue curve for which $D_{\lambda}$ goes positive.

\begin{figure}[t!]
\centering
\includegraphics[trim = 40 130 250 20, width = 1\linewidth]{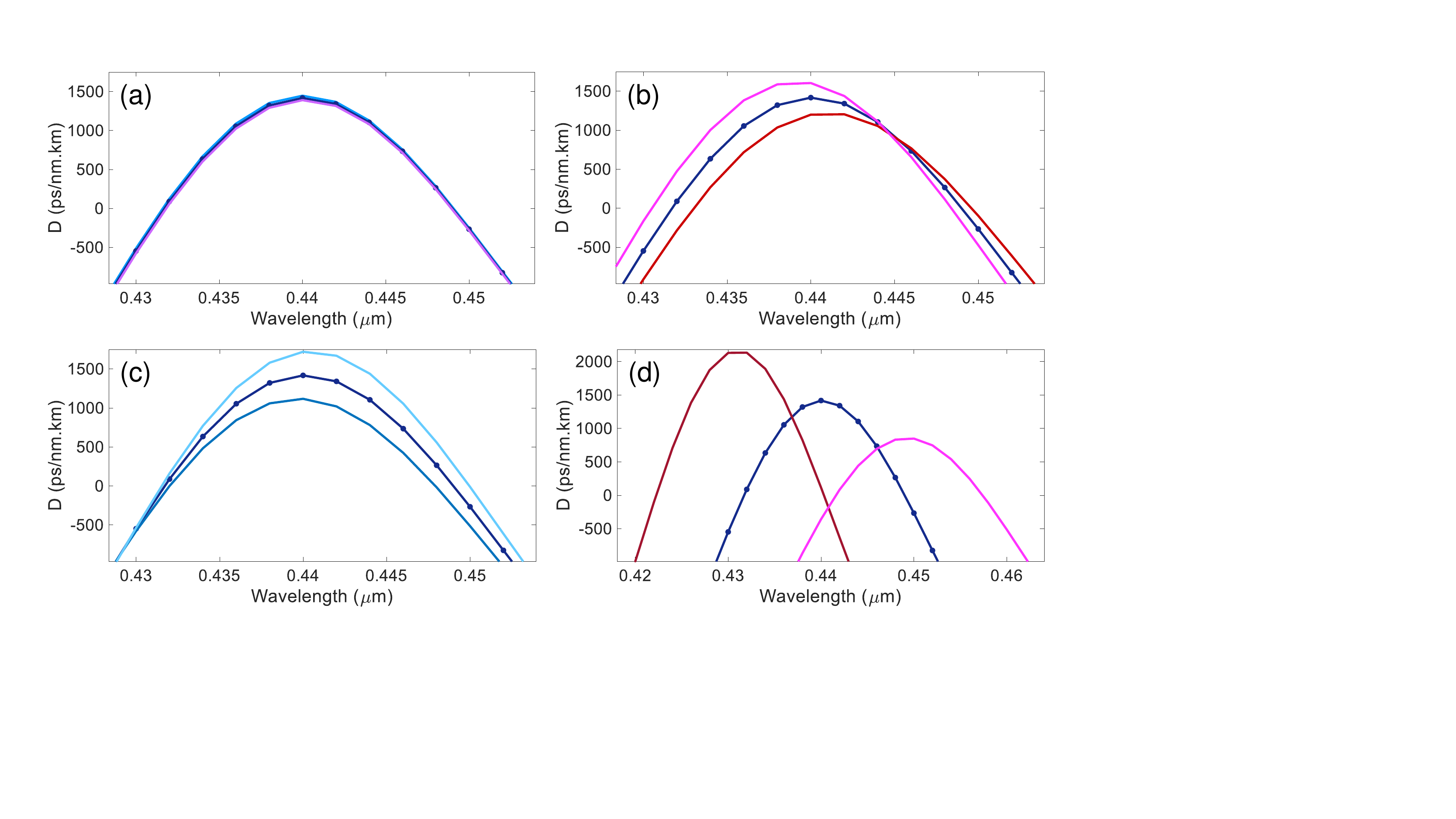}
\caption{Sensitivity of the GVD parameter $D_{\lambda}$ to the geometric parameters of the hybrid structure shown in Fig.~\ref{fig:guide}.  The violet curves with dots correspond to the optimal (underlined) design values.  (a) Guide width $w = 680$ (blue), $\underline{700}$, and $720$~nm (magenta).  (b) Guide 1 height $h_{\textrm{f}1} = 435$ (brown), $\underline{440}$, and $445$~nm (magenta).  (c) Gap height $h_{\textrm{g}} = 245$ (blue), $\underline{250}$, and $255$~nm (cyan).  (d) Guide 2 height $h_{\textrm{f}2} = 195$ (brown), $\underline{200}$, and $205$~nm (magenta).}
\label{fig:dispersion}
\end{figure}

We next assessed the sensitivity of the GVD parameter to the key geometric parameters of the hybrid structure, namely, its width, the gap height, and heights of the individual guides.  The results of this study are summarized in Figure~\ref{fig:dispersion}, where the curves with dots indicate the GVD profile of the optimal design shown in Figure~\ref{fig:guide}; \emph{i.e.}, they are the same as the solid blue $D_{\lambda}$ curve shown in Fig.~\ref{fig:guide}(b).

Deviations of the waveguide width $w$ from its target value can arise from imprecise etching of the epitaxial layers.  Accordingly, we investigated a representative range of $\pm20$~nm around the nominal design value of $700$~nm.  As can be seen in Fig.~\ref{fig:dispersion}(a), this has an imperceptible impact on $D_{\lambda}$, which is not surprising since $w$ is appreciably larger than $\lambda$ in all three cases.

Deviations from target values of the height parameters can be expected to be negligible, since the epitaxial layer thickness is usually a well-controlled parameter at the level of a few monolayers of resolution.  This fact notwithstanding, even small deviations from an optimal design value in the vertical direction may be expected to significantly impact the GVD parameter of the resulting hybrid structure, since this is the direction along which crucial coupling takes place.  With this in mind, we investigated a modest $\pm5$~nm range of variation around the target values of $h_{\textrm{f}1}$, $h_{\textrm{f}2}$, and $h_{\textrm{g}}$.  As can be seen in Figs.~\ref{fig:dispersion}(b) and 3(d), increasing $h_{\textrm{f}1}$ results in a gradual shifting of the $D_{\lambda}$ peak up and to the left (in wavelength), while increasing $h_{\textrm{f}2}$ shifts the $D_{\lambda}$ peak much more dramatically and in the opposite direction (down and to the right).  The reason for this is the considerable difference in the slopes of the dispersion curves $\beta_{1}$ and $\beta_{2}$ near the avoided-crossing point, which are, in turn, controlled by the aspect ratios of the individual guides \cite{Chua2009}.  Finally, an increase in $h_{\textrm{g}}$ causes little change in the position of the peak but does enhance the value of $D_{\lambda}$ significantly, as seen in Fig.~\ref{fig:dispersion}(c).  This can be understood from our results in the previous section, as follows.  The coupling coefficient $\kappa$ is given approximately by an overlap integral between the modes of the individual guides \cite{Chua2009}.  Consequently, a larger $h_{\textrm{g}}$ translates into a smaller $\kappa$, which, in turn, makes the second term in \eqref{eq:betappwc} larger, with the overall effect of driving $\beta''$ more negative.

As mentioned before, the run-to-run processing variability in these geometric parameters is likely to be smaller than the $\pm5$~nm range considered here.  The observed sensitivity is therefore more relevant as a demonstration of the range of tunability of the anomalous dispersion response.  By suitably optimizing the thicknesses of the high-index regions as well as the thickness of the gap, it should be possible to increase or decrease the peak $D_{\lambda}$ value while shifting the positive-$D_{\lambda}$ band toward shorter or longer wavelengths, thus tailoring the anomalous dispersion response of the structure as desired.

\section{Comb generation}
With anomalous dispersion in the blue/UV part of the spectrum achievable in a III-Nitride material system through advanced dispersion engineering as described in the previous section, we are motivated to investigate the formation of a Kerr cavity soliton in a microring resonator with radius $R$ and transverse cross-section as shown in Fig.~\ref{fig:guide}.  Closing the waveguide upon itself in this manner results in a (countably infinite) set of cavity modes whose resonant frequencies satisfy $2 \pi R \, \beta(\omega) = 2 \pi l$; \emph{i.e.}, they are given by
\begin{equation}
\omega_{l} = \zeta\left(\frac{l}{R}\right)
\label{eq:omegaell}
\end{equation}
for positive integer $l$, where $\zeta(\cdot) = \beta^{-1}(\cdot)$.\footnote{It should perhaps be mentioned that one must apply a ``curvature'' correction (of order $w/R$) to $\beta(\omega)$ of the linear waveguide in order to obtain the corresponding propagation constant of the circular waveguide.  [It is this latter function that enters into \eqref{eq:omegaell} and \eqref{eq:LLE}].  For the geometric parameters chosen here, this is indeed a small correction; however, it should be accounted for when considering the optimal placement of a linear waveguide section for coupling into and out of the circular resonator.}

Assume now that the resonator is pumped by a continuous-wave (CW) laser with frequency $\omega_{0}$ that is near a cavity resonance at $\omega_{\ell}$.  Let $\phi$ denote the (azimuthal) coordinate around the resonator, and transform to a frame that is rotating with the group velocity (at $\omega_{\ell}$) in the cavity by defining $\theta = \phi - (v_{\textrm{g}}/R) \, t$.  The spatio-temporal evolution\footnote{This is essentially a mean-field approximation to the more rigorous coupled-mode approach \cite{chembo-cmt}.  The latter provides an alternative spectro-temporal description of the process, which is quite a bit more costly to model computationally.} of the normalized intra-cavity mode amplitude $\psi(\theta, t)$ is then found to be governed by the Lugiato--Lefever equation (LLE) \cite{chembo-lle}
\begin{equation}
\frac{\partial \psi}{\partial t} = -(1 + \mathrm{i} \tilde{\delta}) \psi + \mathrm{i} |\psi|^{2} \psi - \mathrm{i} \, \frac{\tilde{\beta}}{2} \, \frac{\partial^{2} \psi}{\partial \theta^{2}} + F,
\label{eq:LLE}
\end{equation}
where $F$ is the normalized pump amplitude in the cavity, $\tilde{\delta}$ is proportional to the detuning $\delta = \omega_{0} - \omega_{\ell}$ of the pump frequency from the nearest cold-cavity resonance, and $\tilde{\beta}$ is proportional to the GVD parameter $\beta''$ and therefore indicates the nature and the strength of dispersion in the cavity.  The LLE is a driven nonlinear Schr{\" o}dinger equation with damping\footnote{Cavity losses include (small) material absorption, radiation loss that is exacerbated by the bending of the guide, and (small) transmittance of the coupling port.  An overall damping coeficient $\alpha$ does not appear explicitly in \eqref{eq:LLE}, however, since it is aggregated with other terms during the normalization process.} and detuning, and therefore soliton formation may be expected for $\tilde{\beta} < 0$.

In addition to anomalous dispersion, another important consideration for Kerr comb generation is the competition between four-wave mixing gain and Raman gain.  The former is the very mechanism that leads to comb formation while the latter tends to be stronger, and therefore it is crucial to ensure their separation in the frequency domain.  For a laser wavelength of $\lambda_0 = 442$~nm, the anti-Stokes lines in crystalline AlGaN are separated by about $18$~THz from the pump frequency, with a linewidth of about $100$~GHz \cite{JGL$^{+}$2019}.  By choosing the ring radius to be $R = 40~\mu$m, we place the free spectral range of the resonator at about $400$~GHz, thus minimizing the overlap between the Raman gain spectrum and the cavity modes.

\begin{figure}[t!]
\centering
\includegraphics[trim = 60 60 90 20, width = 1\linewidth]{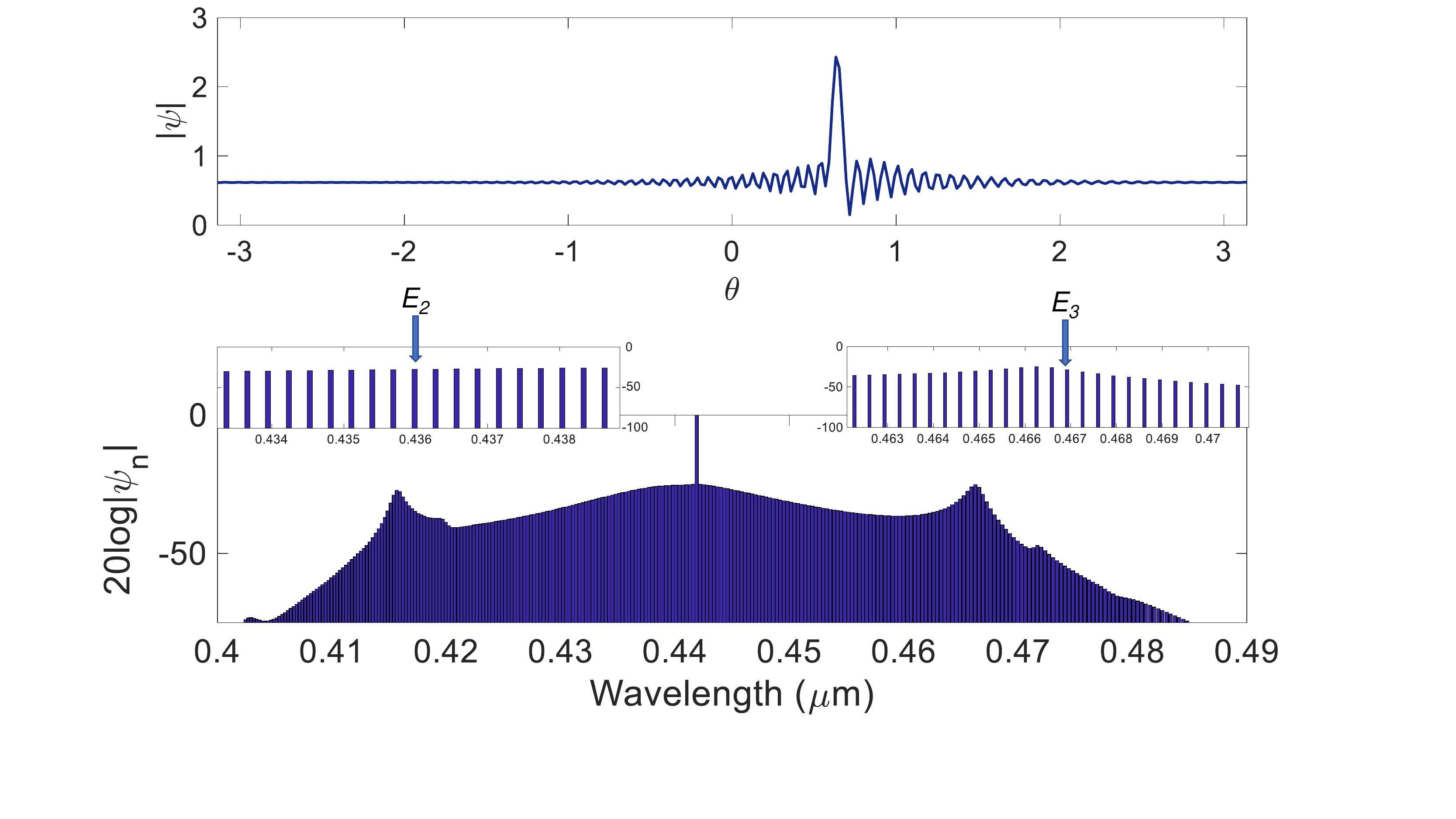}
\caption{Intra-cavity Kerr soliton (top) and associated normalized frequency-comb spectrum (bottom) in a microring resonator with radius $R = 40~\mu$m and transverse cross-section shown in Fig.~\ref{fig:guide}, driven by a CW laser at $\lambda_0 = 442$~nm.  The two insets highlight the specific comb lines matching the $^{171}{\textrm{Yb}}^{+}$ ion optical clock transitions at $436$~nm ($E_2$) and $467$~nm ($E_3$), respectively.} 
\label{fig:comb}
\end{figure}

With a normalized detuning of $\tilde{\delta} = 3.5$ and a normalized pump power of $F^2 = 3.85$, the numerical solution of \eqref{eq:LLE} via the standard split-step Fourier method \cite{WH1986} yields the bright Kerr soliton and the corresponding frequency-comb spectrum shown in Figure~\ref{fig:comb}.  As can be seen, the comb spans a wide spectral range in excess of $75$~nm at the $-70$~dB window, thus providing strong comb lines at the optical clock transitions $E_2$ and $E_3$ of $^{171}{\textrm{Yb}}^{+}$.

\section{Conclusion}
In this paper, our goal was to leverage the state-of-the-art photonic materials and fabrication techniques toward designing an integrated optical frequency combs in the UV and short-wavelength-visible regions of the spectrum.  With the relevant transitions of the $^{171}{\textrm{Yb}}^{+}$ ion targeted for concreteness in the design exercise reported here, we demonstrated the design of an anomalously dispersive hybrid micro-resonator structure using III-Nitride family materials that exhibit normal dispersion in this spectral regime.  The design approach exploits the avoided-crossing phenomenon in coupled waveguides, and leads to practicably realizable resonator dimensions while allowing for a wide range of tunability of the dispersion response.  Numerical solutions of the Lugiato--Lefever equation demonstrated the capability of the designed micro-resonator to support the formation of a broadband Kerr-soliton blue/UV frequency comb under CW pumping.



\end{document}